\documentclass[aps,prb,twocolumn,groupedaddress]{revtex4}

\usepackage{graphicx}
\usepackage{dcolumn}
\usepackage{bm}
\usepackage{longtable}
\usepackage{subfigure}
\usepackage{amsmath}

\newcommand{\kj}[1]{$\rm{#1}$}

\begin{document}


\title{Improved Tight Binding Parametrization for the
Simulation of Stacking Faults in Aluminium}


\author{Anders G. Fr\o seth}
\email[Email: ]{Anders.Froseth@phys.ntnu.no}
\affiliation{Department of Physics,
Norwegian University of Science and Technology (NTNU), N-7491 Trondheim, Norway}
\author{Peter M. Derlet}
\affiliation{Paul Scherrer Institute, CH-5232 Villigen PSI, Switzerland}
\author{Randi Holmestad}
\affiliation{Department of Physics,
Norwegian University of Science and Technology (NTNU), N-7491 Trondheim, Norway}
\author{Knut Marthinsen}
\affiliation{Department of Materials Technology,
Norwegian University of Science and Technology (NTNU), N-7491 Trondheim, Norway}


\date{\today}

\begin{abstract}
We refit the NRL tight binding parameterization for Aluminium by Mehl 
\emph{et al} [Phys. Rev. B, 61, 4894 (2000)], to a database generated 
via full potential Linearized Augmented Plane Wave (LAPW) Density 
Functional Theory (DFT) calculations. This is performed using a global 
optimization algorithm paying particular attention to reproducing the 
correct order of the angular symmetries of the tight binding fcc and 
bcc bandstructure. The resulting parameterization is found to better 
predict the hcp phase and both the stable and unstable planar stacking 
fault defect energies.
\end{abstract}


\maketitle


The empirical tight binding technique can be seen as a compromise
between the very accurate but computationally expensive \emph{ab
initio} Density Functional Theory (DFT) methods, and the fast but
less accurate empirical potential methods such as the Embedded
Atom Method (EAM). Although great reductions in computational cost
has been achieved for \emph{ab initio} methods by employing a
minimal basis set and efficient algorithms that scale linearly
with atom number \cite{siesta}, fully self-consistent calculations
are still relatively slow for ``large'' scale Molecular Dynamics
(MD) simulations. For empirical tight binding methods, a range of
linear scaling techniques have also been developed
\cite{lin_TB_comp} and the development of accurate tight binding
parameterizations is of utmost importance, if this method is to
remain competitive. Furthermore, the strength of the
tight binding method, contrary to many EAM parameterizations, is
its ability to accurately predict a range of structural properties
not included in the fitting database.

The NRL tight binding formalism is a total energy tight binding
method based on the fitting of bandstructure eigenvalues and total
energies from \emph{ab initio} calculations. One of the advantages
of this method is that one is able to fit to the bandstructure and
total energy of a given structure simultaneously be redefining the
energy zero of the sum of eigenvalues. From DFT, the total energy
can be expressed as the sum of the single electron eigenvalues
plus an ionic-like term
\begin{equation}
E(n(\mathbf{r})) = \sum_if(\mu-\epsilon_i)\epsilon_i + F(n(\mathbf{r}))
\end{equation}
where $f(x)$ is the Fermi function, $\epsilon_i$ the electronic
eigenvalues, $\mu$ the chemical potential, and $F(n(\mathbf{r}))$
the so-called ``correction term'' \cite{NRL_TB,NRL_TB_Al}.
Following Mehl \emph{et al} this can be reformulated by shifting the
energy zero of the bandstructure term by a value exactly
canceling the ``correction term'':
\begin{equation}
E(n(\mathbf{r})) = \sum_if(\mu'-\epsilon'_i)\epsilon'_i
\end{equation}
This shift is taken into account by making the onsite terms
environmentally dependent. Using a nonorthogonal two-center
Slater-Koster formulation \cite{SK}, the tight binding model is
then parameterized by fitting the bandstructure and total energies
as a function of lattice constant and crystal structure of the
tight binding model to a database produced by \emph{ab inito}
calculations. The philosophy of the NRL formalism is to include
only a handful of high symmetry structures in the fitting
database. This method has proven to be successful for a wide range
of elements. Further details can be found in Refs. \cite{NRL_TB,
Mehl_review}.

We have refitted a tight binding parameterization by Mehl \emph{et
al} \cite{NRL_Rice} for Aluminium using an $s$, $p$ and $d$
orbital basis, to the full potential LAPW DFT calculations of
WIEN2k\cite{wien2k}. As an alternative to the conventional least
squares procedure we have used Adaptive Simulated Annealing (ASA),
a stochastic global optimization algorithm \cite{My_TB,ASA}.
Assuming a probability density of states $g_T(\mathbf{x})$ for the
parameters space $\{x_i\}$, and a cost function defined by the
``energy'' $E(x)$. At any step, $k$, the ``probability'' for
accepting the new cost function $E_{k+1}$ is given by:
\begin{equation}
h(\Delta E) =
\begin{cases}
\exp(-\Delta E/T)& \text{for $\Delta E>0$}, \\
1& \text{for $\Delta E<0$}.
\end{cases}
\end{equation}
where the parameter $T=T(k)$ defines the ``annealing schedule''.
For such a system the principle of detailed balance holds, which
means that, in theory, all states of the system will be sampled by
the algorithm and a global minimum can be found. A weakness of the
simulated annealing approach is that it needs to sample a large
part of parameter space and can therefore be inherently slow. This
is partly remedied in the ASA algorithm by assuming a probability
density for the parameters with ``fat'' logarithmic tails
\begin{equation}
g_T(x) \equiv \prod_i g^i_T(x_i) = \prod_i \frac{1}{2(|x_i|+T_i)\ln(1+T_i)}
\end{equation}
where $x=\{x_i \in [-1,1];i=1,N\}$ is the normalized parameter
set. This allows for a very fast annealing schedule
\begin{equation}
T_i(k) = T_{t0}\exp{-c_ik^{1/N}}
\end{equation}
in contrast to the conventional linearly decreasing annealing
schedule. $T_{i0}$ and $c_i$ are free parameters, which allows the
algorithm to adapt to a changing ``environment'' for the cost
functions as the search proceeds. Further details can be found in
ref. \onlinecite{My_TB}.

In the fitting database we included band-structure data as a
function of lattice constant for the the fcc and bcc crystal
structures using irreducible k-point meshes of 146 for fcc and 91
points for bcc. In order to ensure the correct angular character
for each eigenvalue and further constrain the Slater-Koster
parameters we block-diagonalized the Hamiltonian for a large
number of high symmetry points and directions: the $\Gamma$, $X$,
$L$, $W$ points and $\Lambda$, $\Delta$, $\Sigma$, $S$, $Z$
directions for the fcc structure, and the $\Gamma$, $H$, $P$, $N$
points and $\Lambda$, $\Delta$, $\Sigma$, $F$, $D$, $G$ directions
for the bcc structure.

For the DFT calculations we used the input parameters:
$R_{mt}=2.5$ Bohr, $R_{mt}K_{max} = 7$ and $G_{max} = 14$
\kj{Ry^{1/2}}. Here, $R_{mt}$ is the muffin tin radius, $K_{max}$
is the plane wave cut-off and $G_{max}$ is the maximum Fourier
component of the electron density. For the exchange-correlation
potential we used the Generalized Gradient Approximation (GGA) of
Perdew et. al \cite{GGA}.

\begin{figure}
\begin{tabular}{c}
\mbox{a)}{\rotatebox{-90.00}{\scalebox{0.30}{\includegraphics{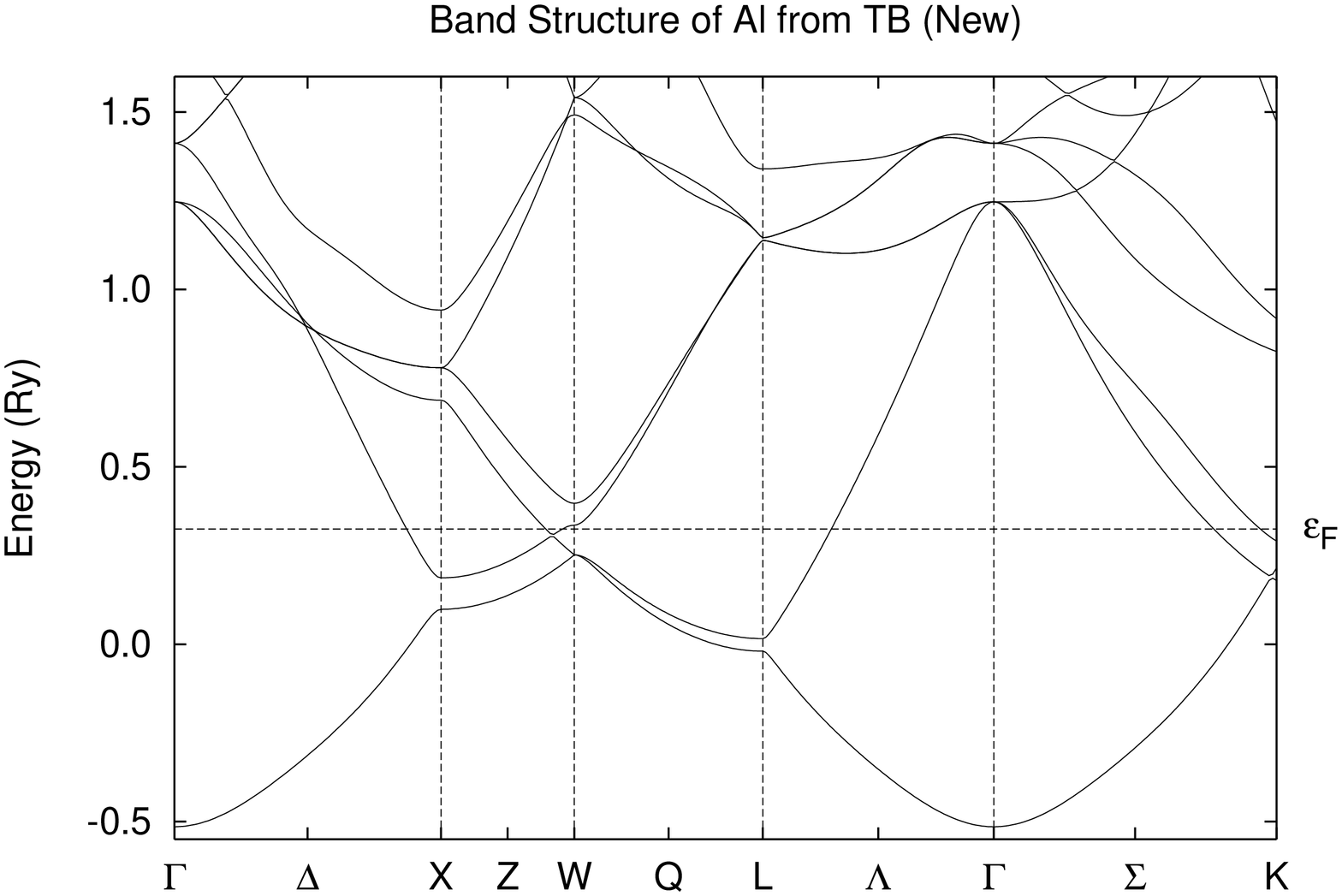}}} }\\
\mbox{b)}{\rotatebox{-90.00}{\scalebox{0.30}{\includegraphics{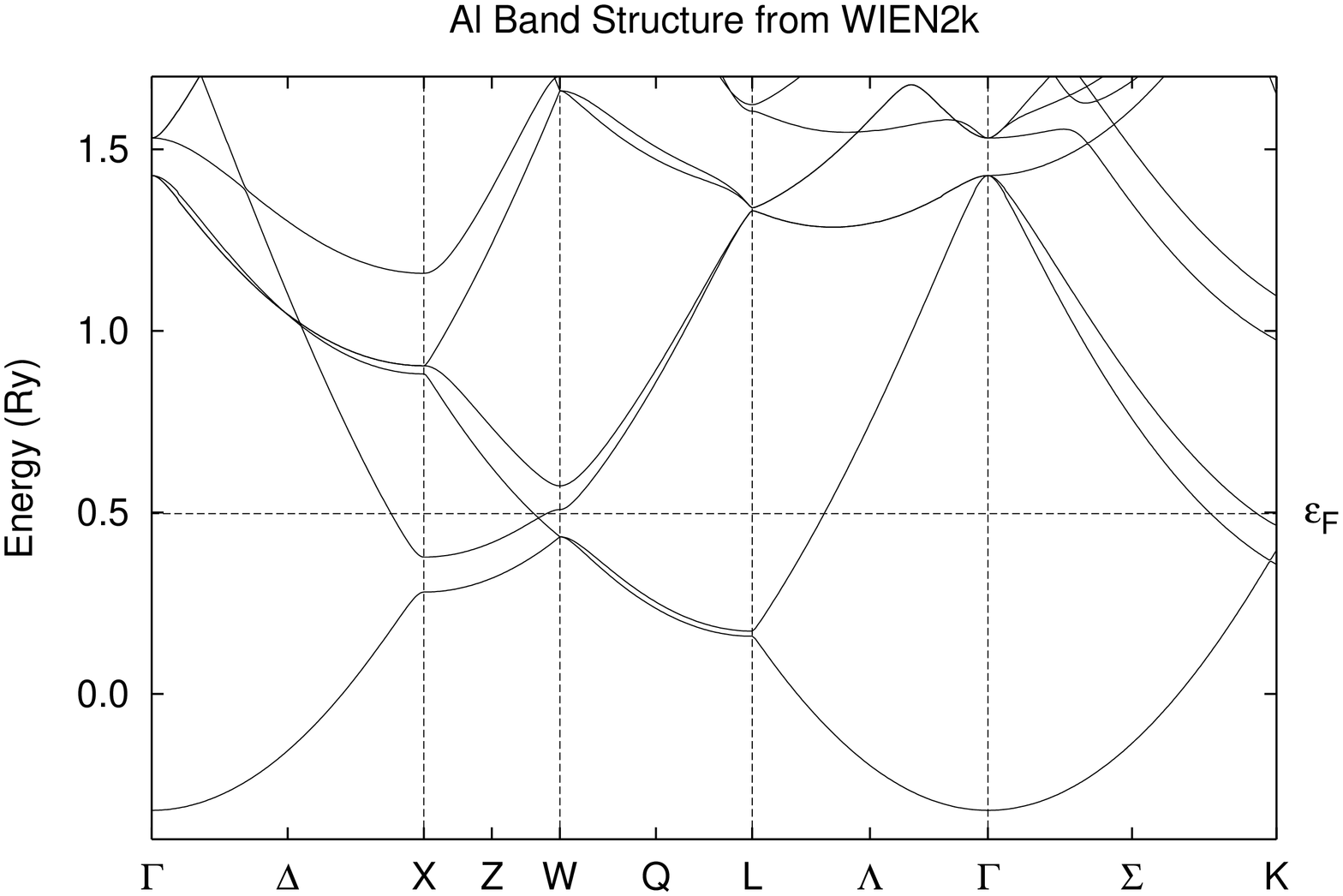}}} }\\
\mbox{c)}{\rotatebox{-90.00}{\scalebox{0.30}{\includegraphics{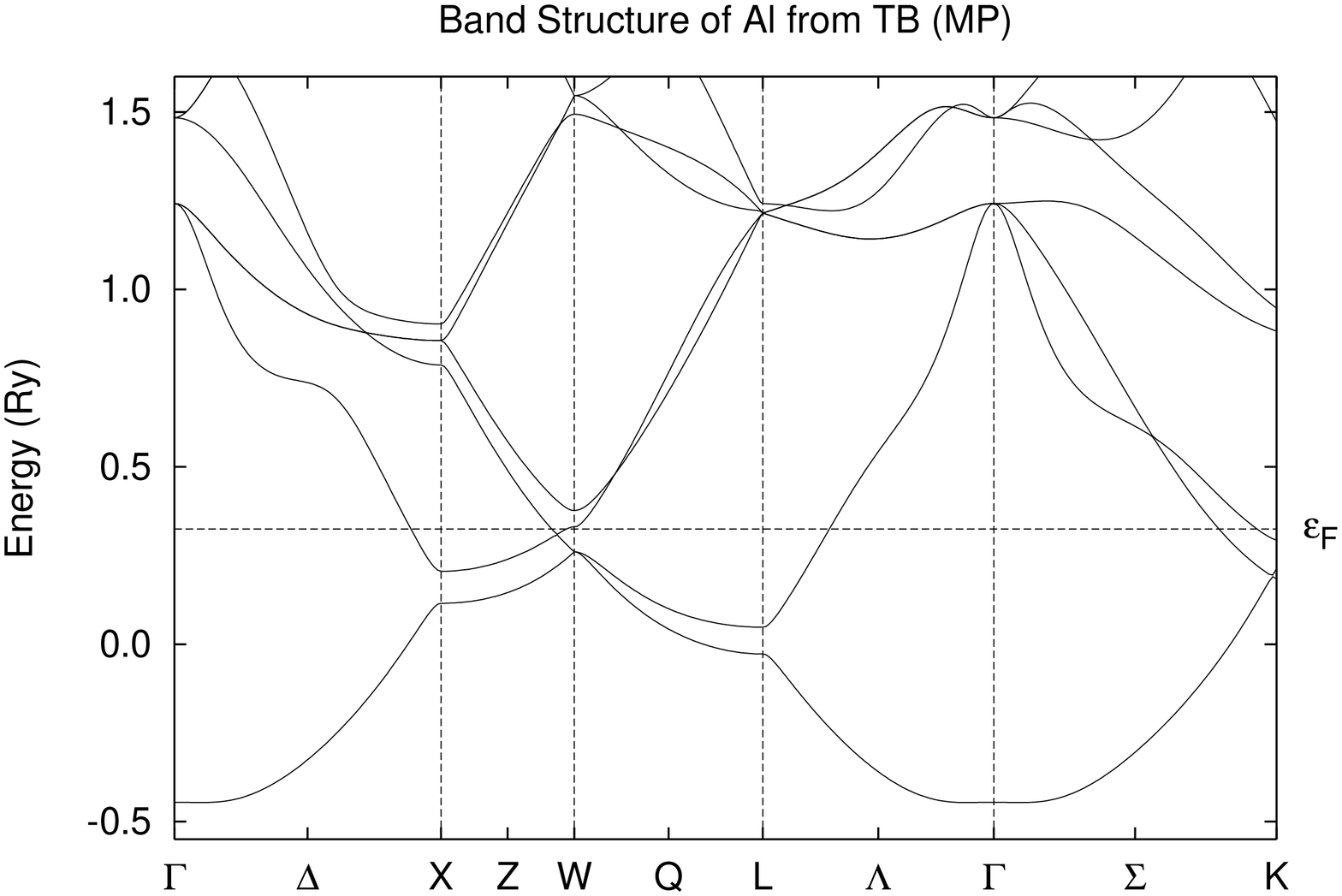}}} }
\end{tabular}
\caption{\label{bands} Band structure for fcc Al at $a=7.65$ Bohr
for a) the new TB parameterization, b) WIEN2k \emph{ab initio}
calculations, and c) the TB parameterization of Mehl et. al.}
\end{figure}

Fig.~\ref{bands} displays the TB bandstructure for fcc at $a=7.65$
Bohr compared to FLAPW \emph{ab inito} calculations and the TB
parameterization of Mehl \emph{et. al}. In contrast to the
previous parametrization, the ordering with respect to angular character of
all bands below 1.25 Ry is now correct. This is important, since
otherwise the relative contribution to the bandstructure from each
atomic orbital in the LCAO-expansion would break the basic
symmetry-requirements of the crystal potential
\cite{group_theory}.

\begin{figure}
\rotatebox{-90.00}{\scalebox{0.30}{\includegraphics{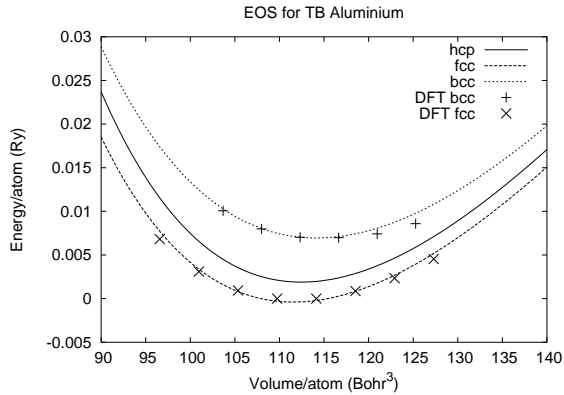}}}
\caption{\label{EOS} Energies of state for the bcc, fcc, and hcp phases.
Note that only information from bcc and fcc structures was included in the fitting database.}
\end{figure}

Fig.~\ref{EOS} displays the energies of state for the fcc, hcp and
bcc structures as obtained from a second order Birch fit to the
data points derived from the present TB model. Also shown are
FLAPW \emph{ab inito} data for fcc and bcc.
Tab.~\ref{elastic_constants} compares the elastic constants from
the TB parameterization of the present work with the
parameterization of Mehl \emph{et al}\cite{NRL_TB_Al}, \emph{ab initio}
DFT calculations using WIEN2k and experimental values. As can be
seen, the TB values are in excellent agreement with the \emph{ab
inito} and experimental values, where only the bulk modulus is
overestimated. This later discrepancy is probably due to more 
weight being given (in the present fitting process) to bandstructure 
data than the equation of state total energy data. The hcp data is in
effect a prediction of the tight binding parameterization since
only the fcc and bcc structures where fitted. At the obtained
equilibrium lattice constants, our fit predicts a
fcc-hcp energy difference of 2.1 mRyd per atom, which agrees well with that
of the converged (with respect to k-point mesh density using the
previously stated \emph{ab initio} parameters) WIEN2k value of 2.0
mRyd. The TB parametrization of Mehl \emph{et. al.} returns a value of 
1.8 mRyd.

\begin{table}
\caption{\label{elastic_constants} Elastic constants for fcc in units
of GPa calculated at $a=7.65$ Bohr. Experimental results are taken from ref. \onlinecite{Handbook}}
\begin{ruledtabular}
\begin{tabular}{lcccr}
Property & TB & TB MP & DFT & Exp. \\
\hline
$a$ & 7.64 & 7.58 & 7.65 & 7.65 \\
$B$ & 92 & 79 & 78 & 79 \\
$C_{11}-C_{12}$ & 48 & 59 & 50 & 46 \\
$C_{44}$ & 27 & 23 & 27 & 28 \\
\end{tabular}
\end{ruledtabular}
\end{table}

\begin{figure}
\rotatebox{-90.00}{\scalebox{0.30}{\includegraphics{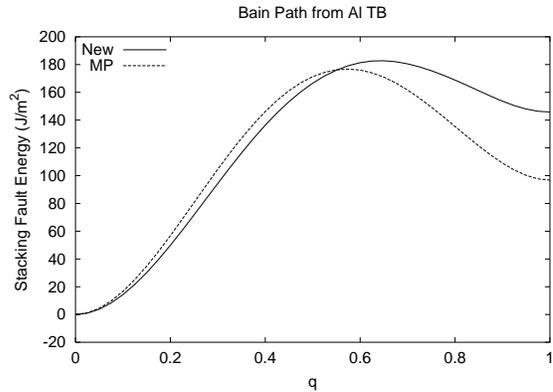}}}
\caption{\label{bain} Bain Path at $a=7.65$ Bohr comparing the new
parameterization with that of Mehl et. al (MP).}
\end{figure}

An important fcc material property for any interatomic model to
reproduce is the intrinsic stacking fault surface energy density
since it plays a crucial role in the disassociation width of a
full dislocation into its constituent Shockley partials and is
believed to play a role in the issue of why in some cases only
partial dislocations are observed in atomistic modeling of the
mechanical properties of nanocrystalline materials
\cite{Yamakov,VanSwygenhoven}. An unrelaxed stacking fault is
modeled by stacking layers of closed packed planes in the [111]
direction using a particular stacking sequence. We choose the
primitive vectors of the unit cell to be:
\begin{equation}
\begin{split}
\mathbf{a_1} = & \frac{1}{2}a \mathbf{y} + \frac{1}{2}a \mathbf{z} \\
\mathbf{a_2} = & \frac{1}{2}a \mathbf{x} + \frac{1}{2}a \mathbf{z} \\
\mathbf{a_3} = & (4+\frac{q}{6})a \mathbf{x} + (4+\frac{q}{6})a \mathbf{y} - (4-\frac{q}{3})a \mathbf{z}
\end{split}
\end{equation}
where $a$ is the lattice constant. $q$ is a variable controlling
the slip in the [112] direction. This is equivalent to a supercell
of 12 close-packed planes, with a variable slip at the 12th layer
due to the periodic boundary conditions. For a $q=0$ we get a
perfect fcc crystal represented by the stacking sequence $ABCABC$,
and for $q=1$ we get the full stacking fault $ABC|BC$, where the
stacking sequence at the interface corresponds to two 111 planes
in which the atoms are hcp coordinated. A mapping of the
structural energy for all values of $q$ from $0$ to $1$ is called
a \emph{Bain Path} (or also a generalized stacking fault curve)
where an intrinsic staking fault occurs at $q=1$. Fig.~\ref{bain}
shows the stacking fault energy as a function of the slip
variable, $q$ relative to the fcc energy at $q=0$. The maximum of
the curve defines the unstable stacking fault configuration, which
in fig.~\ref{bain} is near $q=0.6$.

Table \ref{sf_energies} presents the calculated values for the
intrinsic stacking fault energy, $\gamma_{is}$, unstable stacking
fault energy, $\gamma_{us}$, and the ratio
$\gamma_{is}/\gamma_{us}$, compared to the TB parameterization of
Mehl et. al\cite{Mehl_Rice-criterion}. In addition we include
published results from DFT and EAM calculations, and experimental
values. The experimentally derived values for $\gamma_{is}$ vary
from a low value of 110 \kj{mJ/m^2} to a high value of 280
\kj{mJ/m^2}. However, \emph{ab initio} DFT calculations have
consistently shown results in the range 161-166 \kj{mJ/m^2}. The
EAM potential of Voter and Chen shows values for the stacking
fault energies that are seriously underestimated. The important
point here is that these values were not included in the fitting
database for this potential \cite{VoterChen}. Similar results have
been produced by the EAM potential of Ercolessi and Adams, which
has been used in large scale Molecular Dynamics (MD) simulations
of extended dislocations \cite{Yamakov}. In contrast to this, the
recently published EAM potential of Mishin and Farkas, show values
for the intrinsic and unstable stacking fault energies which are
more in line with the \emph{ab initio} calculations. The reason
for this is that a value for the intrinsic stacking fault energy
was included in the fitting database \cite{MishinFarkas}. The TB
parameterization of the present work shows a marked improvement
over the older parameterization of Mehl \emph{et al} Especially the
intrinsic stacking fault energy, $\gamma_{is}$, with a value of
146 \kj{mJ/m^2} which is in much better agreement with \emph{ab
inito} calculations than a value of 97 \kj{mJ/m^2} of the original
fit. Another important improvement is the value of the ratio
$\gamma_{us}/\gamma_{is}$, which is a measure of the relative cost
of dislocation nucleation with respect to the intrinsic stacking
fault energy. For the new TB parameterization this value is 1.3
which is in good agreement with the value of 1.4 from \emph{ab
inito} calculations.

\begin{table*}
\caption{\label{sf_energies} Unstable and intrinsic stacking
fault energies for Aluminium in \kj{mJ/m^2}}
\begin{ruledtabular}
\begin{tabular}{lcccccr}
Property & TB & TB & \emph{ab initio} & EAM & EAM & Exp.\footnotemark[5] \\
         & Present work & Mehl \emph{et al}\footnotemark[1] & DFT\footnotemark[2]
         & Mishin and Farkas\footnotemark[3]
     & Voter and Chen\footnotemark[4] &   \\
\hline
$\gamma_{us}$ & 183 & 177 & 224 & 167 & 93 & -- \\
$\gamma_{is}$ & 146 & 97 & 166 & 146 & 76 & 110-216 \\
$\gamma_{us}/\gamma_{is}$ & 1.3 & 1.8 & 1.4 & 1.2 & 1.2 & -- \\
\end{tabular}
\end{ruledtabular}
\footnotetext[1]{Using the parameters of ref. \onlinecite{Mehl_Rice-criterion}}
\footnotetext[2]{Ref. \onlinecite{Kaxiras}}
\footnotetext[3]{Ref. \onlinecite{MishinFarkas}}
\footnotetext[4]{Ref. \onlinecite{VoterChen}}
\footnotetext[5]{High value from ref. \onlinecite{SF_exp_high}, low value ref. \onlinecite{SF_exp_low}}
\end{table*}

The improved prediction of the stacking fault energies is
consistent with the corresponding improvement of the hcp cohesive
energy and lattice constant of the present model. At the nearest
neighbor bond length, the local fcc and hcp environments differ
only with respect to their dihedral angle arrangements. In terms
of energy and lattice constant this difference should be sensitive
to the angular character of, particularly, the $\pi$ and $\delta$
bonds of the $p$ and $d$ states, and it is precisely such bonding
elements that the current parameter-set more accurately reproduces
at and near the Fermi level. 

In conclusion, we have refitted the 
tight binding parameters of Mehl \emph{et al} for Aluminium to \emph{ab
initio} FLAPW eigenvalues. We have used a stochastic global
optimization algorithm and an extended database of high symmetry
eigenvalues. As a result we obtain a parameterization with an
improved representation of the angular symmetries for the Slater
Koster parameters. As a consequence we obtain improved predictions
for the stacking fault energies without including this information
explicitly in the fitting database.

Parts of this work has been supported by the Norwegian Research
Council through CPU time on the NOTUR Origin 3800. The static
tight-binding code was provided by M. J. Mehl of the Naval
Research Laboratory under the U.S. Department of Defense CHSSI
program.

\bibliography{Al_SF_04-03-03}

\end{document}